# Mechanically-Exfoliated Stacks of Thin Films of Bi$_2$Te$_3$ Topological Insulators with Enhanced Thermoelectric Performance

V. Goyal, D. Teweldebrhan and A.A. Balandin[*]

Nano-Device Laboratory, Department of Electrical Engineering and Materials Science and Engineering Program, Bourns College of Engineering, University of California - Riverside, Riverside, California 92521 USA

## Abstract

The authors report on "graphene-like" mechanical exfoliation of single-crystal Bi$_2$Te$_3$ films and thermoelectric characterization of the stacks of such films. Thermal conductivity of the resulting "pseudo-superlattices" was measured by the "hot disk" and "laser flash" techniques. The room-temperature in-plane (cross-plane) thermal conductivity of the stacks decreases by a factor of ~2.4 (3.5) as compared to bulk. The thermal conductivity reduction with preserved electrical properties leads to strong increase in the thermoelectric figure of merit. It is suggested that the film thinning to *few-quintuples* and tuning of the Fermi level can help in achieving the topological-insulator surface transport regime with an extraordinary thermoelectric efficiency.

[*] Corresponding author; electronic address: balandin@ee.ucr.edu ; group web-site: http://ndl.ee.ucr.edu

V. Goyal, D. Teweldebrhan and A.A. Balandin, UC Riverside, submitted to *Applied Physics Letters*, 2010

Bismuth Telluride ($Bi_2Te_3$) is a narrow band gap semiconductor, which is well known as one of the best thermoelectric (TE) materials [1]. At room temperature (RT), bulk $Bi_2Te_3$ and related compounds reveal the highest thermoelectric figures of merit $ZT=S^2\sigma T/(K_e+K_l) \sim 1$, where $S$ is the Seebeck coefficient, $\sigma$ is electrical conductivity, $T$ is the absolute temperature, $K_e$ and $K_l$ are the electron and phonon (lattice) contributions to the thermal conductivity. The quantum confinement of charge carriers [2], acoustic phonon – boundary scattering [3] and spatial confinement of acoustic phonons [4] have been utilized for further increase of $ZT$ in $Bi_2Te_3$ quantum structures. The improvement was predicted to result either from confinement-induced increased electron density of states near the Fermi level $E_F$ [2] or reduction of $K_l$ due to the phonon – boundary [3] scattering or modification of the phonon spectrum [4].

Recent developments renewed interest to $Bi_2Te_3$, and opened up a completely different strategy for $ZT$ enhancement in $Bi_2Te_3$ thin films. It was discovered that $Bi_2Te_3$ family of materials are topological insulators (TIs) [5]. TIs are materials with a bulk insulating gap and conducting surface states that are topologically protected against scattering by the time-reversal symmetry [5-6]. It was shown theoretically that $ZT$ can be strongly enhanced in $Bi_2Te_3$ thin-film TIs provided that the Fermi level is tuned to ensure the surface transport regime and the films are thin enough to open a gap in the "Dirac cone" dispersion on the surface [7].

We previously demonstrated that a "graphene-like" procedure can be used to mechanically exfoliate the ultra-thin films of $Bi_2Te_3$ with the thickness down to a single *quintuple* [8]. The quintuples (thickness $W\sim 1$ nm) are structural units in $Bi_2Te_3$ crystal separated from each other by the van der Waals gaps. The mechanically exfoliated few-quintuple films have a number of benefits compared to grown thin films. They are perfectly crystalline and have an essentially infinite potential barrier for electrons and holes. The latter is a drastic difference from $Bi_2Te_3$-based superlattices grown by molecular beam epitaxy or other techniques. The ultimately high potential barriers together with a few-quintuple thickness and possibility of electrical back-gating can ensure the strong quantum confinement for electrons and $E_F$ fine-tuning.

In this letter we report on the thermoelectric properties of stacks of mechanically exfoliated $Bi_2Te_3$ films focusing on a possibility of *ZT* enhancement via reduction of the thermal conductivity. The thermoelectric applications require a sufficient quantity of material, i.e. "bulk", i.e. the single quintuples would hardly be practical. For this reason, we studied the stacks of the exfoliated films, which were put on top of each other and subjected to thermal treatment.

The details of the "graphene-like" process for exfoliating $Bi_2Te_3$ films were reported by us elsewhere [8]. The resulting stacks were composed of the individual films with different thicknesses (from a few nm to μm). The stacks were intentionally made rather thick (up to ~0.5 mm). The material crystallinity and quality were verified with micro-Raman spectroscopy. Figure 1 shows a spectrum of the exfoliated $Bi_2Te_3$ film recorded under 488-nm laser excitation. The observed peaks, $E_g^1(TO)$, $A^1_{1g}(LO)$, $E_g^2(TO)$, and $A^2_{1g}(LO)$, are consistent with literature for crystalline $Bi_2Te_3$ [9]. It is interesting to note an additional peak, identified as $A_{1u}$, which appears as a result of the crystal symmetry breaking in thin films with the thickness below the light penetration depth [10]. It is not present in the bulk $Bi_2Te_3$ crystals. The intensity of $A_{1u}$ peak and the intensity ratio $I(A_{1g}^2)/I(E_g^2)$ can be used for nanometrology of few-quintuple films [10]. The shown spectra correspond to the film with the thickness $W\sim50$ nm. The thickness was cross-checked with the atomic force microscopy. The inset shows a scanning electron microscopy (SEM) image (Philips XL-30 FEG) of two overlapping exfoliated $Bi_2Te_3$ films.

The "graphene-like" exfoliated films were transferred to a substrate and mechanically put on top of each other. The stacks were annealed at ~250 $^0$C to reduce the air gaps between the layers and improve structural stability. Figure 2 shows the cross-sectional SEM image of a stacked sample. One can see that the films of different thicknesses are assembled into a non-periodic "pseudo-superlattice" (also referred to as compositionally modulated structures or "random superlattices"). The non-periodic "superlattices" may have certain benefits for thermoelectric applications owing to flexibility for tuning the phonon transport [11]. The films elemental composition and stoichiometry were verified with the energy dispersive spectroscopy (EDS). The presence of prominent Si and O peaks indicate that the individual films, which make up the stacks, are transparent to the electron beams.

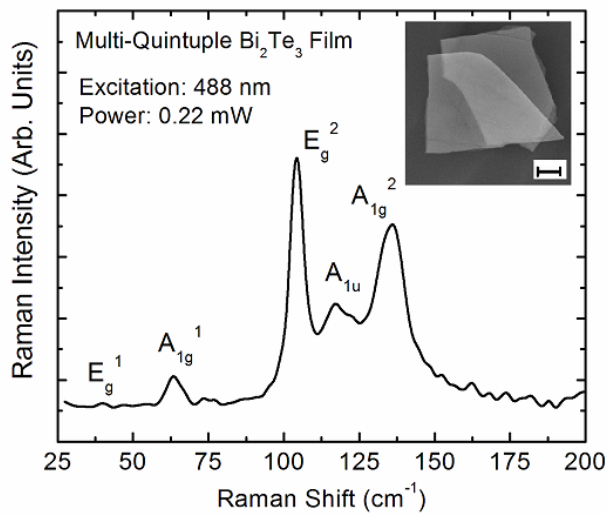

**Figure 1:** Raman spectrum of the "graphene-like" exfoliated $Bi_2Te_3$ films. Note the appearance of $A_{1u}$ peak, not Raman active in bulk crystals, due to the crystal symmetry breaking in thin films. The inset shows the scanning electron microscopy image of two overlapping mechanically exfoliated films. The scale bar in the inset is 1 μm.

The measurements of the thermal conductivity, $K$, were performed by two different experimental techniques to obtain its in-plane and cross-plane components. The first technique was the transient plane source (TPS) "hot disk" technique, which can measure the average in-plane thermal conductivity. The second technique was the optical "laser flash" technique (LFT), which measures the average cross-plane $K$. In TPS technique [12], a short electric pulse is passed through an electrically insulated sensor sandwiched between two pieces of the sample under investigation. The sensor acts both as a heat source and a thermometer to determine the temperature rise, $\Delta T$, in the sample in response to heating. The thermal diffusivity, $\alpha$, and specific heat, required for determining $K$ value, were obtained by measuring $\Delta T$ as a function of time. Our LFT instrument (Netzsch NanoFlash LFA 447) was equipped with a xenon flash lamp which heated the sample from one end by light shots. The temperature rise was determined at the back side of the samples with the nitrogen-cooled InSb IR detector. Using the thermal-wave travel time, we measured $\alpha$ and determined $K$ from the equation $K=\alpha\rho C_p$, where $C_p$ is the heat capacity and $\rho$ is the mass density of the material. The details of both instruments and measurement procedures were reported by us somewhere [13].

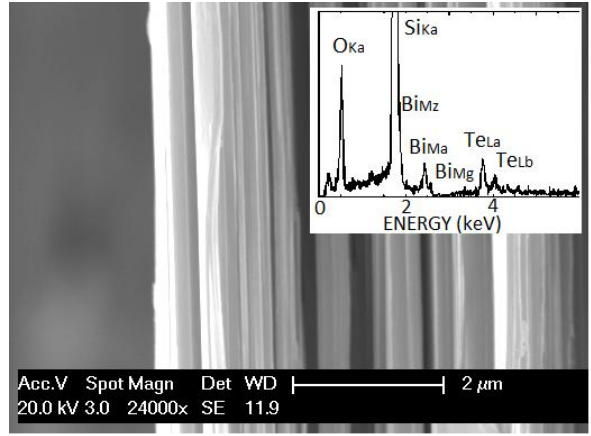

**Figure 2:** SEM image of the stacked "pseudo-superlattice" of the mechanically exfoliated $Bi_2Te_3$ films. The inset shows EDS spectrum of the exfoliated films indicating their atomic composition and transparency for the electron beam.

Figure 3 presents the results of our thermal measurements for three representative "pseudo-superlattices" (samples A with $W\sim0.4$ mm, B with $W\sim0.3$ mm and C with $W\sim0.1$ mm), reference bulk samples, and literature data for bulk $Bi_2Te_3$. Our data for the in-plane $K$ in bulk (denoted with the black-white triangles) are in good agreement with literature values [14] marked by the stars. The monotonic $K$ decrease with $T$ is characteristic for semiconductor materials where thermal transport is limited by the crystal inharmonicity. The data for the bulk cross-plane $K$ are also consistent. The good agreement for bulk attests for accuracy of our measurements. One can see in Figure 3 that there is a strong decrease in the in-plane thermal conductivity of the "pseudo-superlattices" as compared to the bulk. The in-plane $K$ for the stacked $Bi_2Te_3$ superlattices is $\sim0.7$ W/mK, which is a reduction by a factor of $\sim2.4$ from the bulk value of $\sim1.7$ W/mK. The RT cross-plane $K$ of the stacked films is $\sim0.14$ W/mK, which is a significant drop, by a factor of 3.5, from the bulk cross-plane value of ~0.5 W/mK. The "pseudo-superlattice" $K$ is only weakly dependent on temperature. This is expected for the thermal transport limited the phonon – boundary scattering [3, 4, 15]. The cross-plane $K$ reduction is much stronger possibly owing to the large thermal boundary resistance between the exfoliated films. The thermal conduction in our samples was mostly by the acoustic phonons as estimated from the Wiedemann-Franz law. Another observation is that $K$ values obtained for the stacks with different thicknesses are nearly the same. It means that the thermal conductivity was limited by the phonon scattering at the interfaces between the individual $Bi_2Te_3$ layers rather than by the scattering on the outside boundaries of the samples. The overall decrease of $K$ in our stacks is exceptional. It is on the higher end or exceeds that reported for $Bi_2Te_3$ nanoparticles [15], alloy films [16] and highly-textured materials [17]. The cross-plane $K$ in our "pseudo-superlattices" approaches the theoretical *minimum* value predicted for the disordered crystals [18].

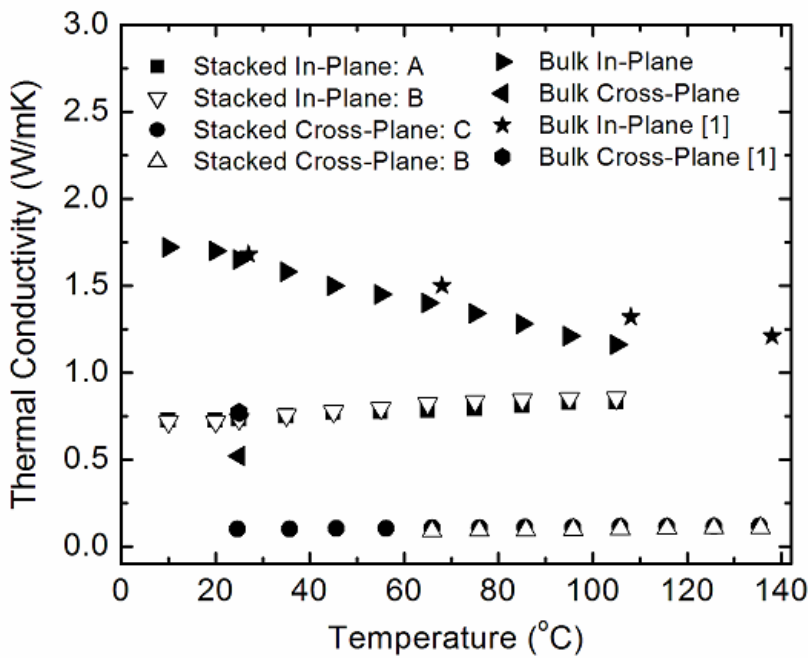

**Figure 3:** Thermal conductivity as a function of temperature for the stacked "pseudo-superlattices" and reference bulk $Bi_2Te_3$ crystals. The literature values for bulk $Bi_2Te_3$ are also shown for comparison.

We analyzed the effect of annealing on $K$. Annealing at 350 $^o$C for 30 seconds led to ~5% increase in the thermal conductivity at RT. The latter was attributed to stronger bonds between the exfoliated $Bi_2Te_3$ layers and further decrease in the air gaps. The annealing at 450 $^o$C for the same time led to the approximately the same decrease in $K$. We explained it by the on-set of inter-diffusion between the layers and increased disorder. The latter was confirmed by the cross-plane SEM studies. The melting point for $Bi_2Te_3$ is rather low ($T_m$~570$^o$C), which supports our conclusion. The Seebeck measurements (MMR SB100) gave $S$ values in the range from 231 to 247 $\mu$V/K. The higher values were obtained for thinner samples. We also investigated the current – voltage characteristics (Signatone and HP4142) of the "pseudo-superlattices" with different thicknesses. The electrical measurements revealed electrical resistivity on the order of ~$10^{-4}$ $\Omega$m, which is close to the optimum for the thermoelectric applications. The strong decrease in the thermal conductivity with preserved electrical properties translates to ~140-250% increase in $ZT$ at RT. It is important to emphasize that the estimated increase is achieved entirely via reduction in the thermal conductivity. The Fermi level in the "pseudo-superlattices" was not optimized. The carrier transport in the films was likely of the mixed volume and surface nature. The further increase in crystal quality, thinning of the films and gating (for achieving the pure surface transport) is expected to result in additional strong $ZT$ increase predicted theoretically [7].

In conclusions, we studied thermoelectric properties of "pseudo-superlattices" prepared by staking of the "graphene-like" mechanically exfoliated of single-crystal $Bi_2Te_3$ films. We showed that $ZT$ in such structures can be substantially increased via reduction of the in-plane and cross-plane thermal conductivity. It is an important observation, since $Bi_2Te_3$ were shown to be topological insulators. Eventually, it may become possible to achieve the pure surface transport regime through the Dirac surface states, topologically protected against scattering, and achieve the theoretically predicted strong enhancement of $ZT$ over a wide temperature range.

*Acknowledgements*

The authors acknowledge the support from SRC – DARPA through the FCRP Center on Functional Engineered Nano Architectonics (FENA).